\documentstyle[12pt]{article}
\newcommand{\oB}{\vert_{\partial M}=0}
\newcommand{\hT}{\tilde h}
\newcommand{\cD}{{\cal D}}
\begin{document}
\begin{titlepage}
\thispagestyle{empty}
\begin{flushright}{TUW 95-06}
\end{flushright}

\centerline{\bf \large On Gauge--Invariant Boundary Conditions
for 2d Gravity}
\centerline{\bf \large with Dynamical Torsion}
\bigskip
\centerline{Dmitri V.Vassilevich \footnote{On leave from:
Department of Theoretical Physics, St.Petersburg
University, 198904 St.Petersburg, Russia. E.mail:
vasilevich@phim.niif.spb.su}}

\vspace{7mm}

\centerline{Institut f\"ur Theoretische Physik}
\centerline{Technische Universit\"at Wien}
\centerline{Wiedner Hauptstr. 8-10, A-1040 Wien}
\centerline{Austria}

\vspace{15mm}

\abstract{In the example of $R^2+T^2$ gravity on the unit
two dimensional disk we demonstrate that in the presence
of an independent spin connection it is possible to define
local gauge invariant boundary conditions even on boundaries
which are {\it not} totally geodesic.
One-loop partition function and the corresponding heat
kernel are calculated.}

\vfill

Vienna, March 1995
\end{titlepage}

{\bf 1.} An important problem in quantum gravity and quantum cosmology
is the formulation of gauge invariant boundary conditions (see
monograph [1] and references therein). This problem is
especially complicated on manifolds with non totally geodesic
boundaries as examplified e.g. by the Euclidean disk. The boundary
conditions used so far in actual computations [2,3,4] either involve
only part of the degrees of freedom [2], or are
only "partially" invariant [3].
The Barvinsky boundary conditions [5], while having all
necessary invariance properties are non-local, which makes
computations very complicated.

Two dimensional quantum gravity is frequently used as a
laboratory for studying various theoretical ideas. As a
particular model we choose here the $R^2+T^2$ gravity [6,7],
which is both classically and quantum integrable [6-8] on manifolds
without boundaries. The aim of this work is twofold. First,
we explore the possibility to define local gauge invariant
boundary conditions in the presence of an independent spin
connection. Second, we study the ultra-violet divergencies
of $R^2+T^2$ gravity due to the presence of boundaries
and comment on the quantum equivalence to a model with only
finite number of quantized modes.

For the sake of simplicity we restrict ourselves to
perturbative one-loop path integral on the background
represented by the unit Euclidean disk. To keep unified
notations with
the Minkowski signature models we call the $O(2)$ rotations
local Lorentz transformations. Our central aim is the definition of
local gauge invariant boundary conditions. Locality means
that the boundary conditions for a field $\Phi$ can be
represented in the form $P_D\Phi \oB$,
$(\partial_0 +c)P_N\Phi \oB$, where $\partial_0$ is the
normal derivative, $P_D$ and $P_N$ are complementary
local projectors. In this context, gauge invariance means
that the gauge
transformations map the functional space defined by
the boundary conditions onto itself. The requirement
of locality seems to be technical. However, it is needed to
obtain a controllable quantum theory. Unlike recent works
[9] on boundary dynamics in dilaton gravity we impose
boundary conditions on all the components of zweibein
fluctuations, i.e. the boundary conditions are imposed
{\it before} fixing a gauge.
This ensures gauge-independence of results.
We discover that local gauge invariant boundary conditions exist
only if the spin connection is independent of the
zweibein. This is exactly the case whenever torsion becomes dynamical.
By computing the heat kernel expansion we demonstrate that
the divergencies proportional to the length of the boundary
are not cancelled. This means that to obtain a model
containing only a finite number of quantized modes
(such models do not give rise to surface terms with
$t^{-\frac 12}$ in the heat kernel expansion) one
should either abandon the locality requirement, or
manifolds with totally geodesic boundaries should be considered.

{\bf 2.} Consider a 2-dimensional diffeomorphism covariant
theory described by zweibein
$e^a_\mu$ and spin-connection
$\omega_\mu^{ab}=$$\omega_\mu \epsilon^{ab}$, $\epsilon^{01}=1$.
For the sake of simplicity let us restrict ourselves to small
fluctuations around the flat two-dimensional Euclidean disk,
then the background
values of zweibein and connection are $e_0^0=1$, $e_1^1=r$,
$\omega_1=1$, where the polar coordinate system is adopted,
$x^1=\theta$, $0 \le \theta \le 2\pi$, $0 \le r \le 1$.
$r=1$ corresponds  to the boundary of the unit disk.

Let us try to define local boundary conditions for the
perturbations $h_\mu^a$ of the zweibein $e_\mu^a$ possessing both
diffeomorphism and Lorentz invariance. The diffeomorphism
transformations with infinitesimal parameter $\xi^\mu$ look
as follows:
$$\delta h_0^0=\partial_0 \xi^0 , \quad
\delta h_0^1=r\partial_0 \xi^1 ,$$
$$\delta h_1^0=\partial_1 \xi^0, \quad
\delta h_1^1=r\partial_1 \xi^1 +\xi^0 . \eqno (1)$$
The boundary conditions for $\xi $ will later become the
boundary conditions for the ghost fields. This is why we
require them to be local too. Let us also adopt the quantum
cosmology boundary conditions for $h_1^1$:
$$h_1^1 \oB \eqno (2)$$
{}From the last of the equations (1) it is clear that $\xi^1$ and
$\xi^0$ should satisfy Dirichlet boundary conditions too:
$$\xi^\mu \oB \eqno (3)$$
{}From (3) we immediately conclude that $h_1^0$ also satisfy the
Dirichlet boundary conditions, while $h_0^0$ and $h_0^1$
obey Neumann bondary conditions with some constants
$c_0$ and $c_1$, whose precise value is irrelevant:
$$h_1^0\oB , \quad (\partial_0+c_0)h_0^0 \oB , \quad
(\partial_0+c_1)h_0^1 \oB \eqno (4)$$
Intuitively it is clear that the normal derivative $\partial_0$
changes the type of boundary conditions. We shall
demonstrate this below in a more rigorous way.

Consider now local Lorentz transformations with infinitesimal
parameter $\sigma$, $\delta h_\mu^a =$$\epsilon^{ab}e_\mu^b
\sigma (x)$. On the disk we obtain
$$\delta h_0^0=\delta h_1^1=0, \quad \delta h_0^1=-\sigma ,
\quad \delta h_1^0=r\sigma . \eqno (5)$$
By comparing the transformation rule (5) for $h_1^0$ with
the boundary condition (4) we are forced to a Dirichlet boundary
condition for $\sigma$
$$\sigma \oB . \eqno (6)$$
However, eq. (6) is in contradiction with the boundary conditions
and the transformation law for $ h_0^1$. We conclude that it is impossible
to define local gauge invariant boundary conditions for the zweibein
fluctuations. A careful analysis shows that this is true for any
manifold with non-vanishing second fundamental form of the boundary
regardless of precise form of the boundary conditions for $ h_1^1$.

Only on a totally geodesic boundary the situation is different.
Then the last of the
equations (1) would contain only the $\xi^1$. This makes it
possible to define local gauge invariant boundary conditions
for the zweibein.

{\bf 3.} To consider the spin-conection $\omega_\mu$ as an
independent field opens completely new possibilities. Indeed,
we may define then a new Lorentz invariant field
$$\tilde e^a_\mu =\exp (-\phi \epsilon^{ab})e^b_\mu , \eqno (7)$$
where $\phi $ is the longitudinal part of the connection
fluctuation $\rho_\mu$ of $\omega_\mu$
$$\rho_\mu =\partial_\mu \phi +\rho_\mu^T, \quad
\nabla^\mu \rho_\mu^T=0. \eqno (8)$$
$\nabla_\mu$ denotes background covariant derivative. The diffeomorphism
transformations of the connection field $\omega$ are
$$\delta \omega_\mu =\xi^\nu \partial_\nu \omega_\mu +
\omega_\nu \partial_\mu \xi^\nu . \eqno (9)$$
Truncated to the linear order in fluctuations over the disk
and transformation parameter they become
$$\delta \phi =\xi^1, \quad \delta \rho^T_\mu =0 . \eqno (10)$$
Thus we see that only the longitudinal part of $\rho$ transforms.
One can easily obtain the linearized transformation rules for
the fluctuations $\hT_\mu^a$ of $\tilde e_\mu^a$
$$\delta \hT_\mu^a=e^{a\nu }\nabla_\mu \xi_\nu , \eqno (11)$$
where $e$ and $\nabla$ are the background zweibein and covariant
derivative, respectively. Repeating step by step our previous
calculations we obtain Dirichlet boundary conditions (3)
for $\xi_\mu$ and mixed boundary conditions for $\hT$
$$\hT^a_1 \oB , \quad (\partial_0 +\frac 1r )\hT^a_0,
\quad a=0,1. \eqno (12)$$
Now the precise form of the Neumann boundary condition is
essential. Consider the derivation of eq. (12) in more
detail. We shall use the observation [10] that it is enough
to define boundary conditions for eigenfunctions of
the Laplace operator $\Delta$. For example,
$$(\partial_0+\frac 1r ) \delta \hT^0_0 =
(\partial_0+\frac 1r ) \partial_0 \xi_0=
(\Delta -\frac 1{r^2} \partial_1^2 +\frac 1{r^2} ) \xi_0
+\frac 2{r^3} \partial_1 \xi_1 , \eqno (13)$$
where we used an explicit expression for the vector Laplace
operator on the disk (see e.g. ref. [1]). The $\xi_\mu$
can be choosen as eigenvector of both $\Delta$ and
$\partial_1$. In this case the r.h.s. of (13) vanishes
term by term on the boundary provided $\xi$ satisfies
the Dirichlet boundary conditions (3).

Using the transformation rule (10) we can now define the
boundary condition for $\phi$
$$\phi \oB . \eqno (14)$$
The equation (14) fixes the boundary condition for the
rotation parameter $\sigma$ and for $\rho_\mu$ since we
require locality of the boundary conditions for the latter
field as well:
$$\sigma \oB , \quad \rho_1 \oB ,\quad \partial_0 \rho_0 \oB .
\eqno (15)$$
In the theory of the de Rham complex the conditions for $\rho_\mu$
(15) are know as relative boundary conditions [11].

Hence in the case of an independent spin-connection we are indeed able
to define local gauge invariant boundary conditions for
the fluctuations $\rho$ and $\hT$.

{\bf 4.} As an example, we
consider the Euclidean $R^2+T^2$ action with zero cosmological
constant.
$$S=\int d^2x e (4R^2+\alpha T^aT^a),$$
$$eR=\epsilon^{\mu \nu} \partial_\mu \omega_\nu , \quad
eT^a=\epsilon^{\mu \nu} (\partial_\mu e_\nu^a-
\omega_\mu \epsilon^{ab} e_\nu^b ) \eqno (16)$$
The background in question is a stationary point of the action (16).

The boundary conditions (12), (15) have the following easily
established properties.
\newline
(i) $\hT$ and $\rho$ admit decompositions
$$\hT^a_\mu =e \epsilon_{\mu \nu} \nabla^\nu v^a+e^{\nu a}
\nabla_\mu \xi_\nu , \quad
\rho_\mu =\partial_\mu \phi +e \epsilon_{\mu \nu} \nabla^\nu s
\eqno (17)$$
with background covariant derivative $\nabla$ and zweibein
$e_\mu^a$.
\newline
(ii) The decompositions (17) are orthogonal with respect
to ordinary inner products without surface terms.
\newline
(iii) The fields $v^a$ and $s$ satisfy Neumann boundary conditions
$$\partial_0 v^a \oB , \quad \partial_0 s \oB . \eqno (18)$$
(iv) The kernel of the map $\{ v^a, \xi_\nu \} \to \hT^a_\mu$
consists of two covariantly constant vectors $v^a$.  The kernel
of the map $\{ \phi ,s\} \to \rho_\mu$ consists of one constant
scalar $s$.

The natural gauge fixing conditions are
$$\nabla^{\mu} \hT^a_\mu =0, \quad \nabla^\mu \rho_\mu =0
\eqno (19)$$
Due to the flatness of the background equations (19) are
equivalent to $\xi =0$ and $\phi =0$.

Now we are able to write down the one loop path integral
in the gauge (19)
$$Z=\int \cD v^a \cD s J_v J_s \exp \left ( -\int d^2xe
[\alpha (\epsilon^{ba}\Delta v^a+\nabla^bs)^2+
s\Delta^2s] \right )$$
$$J_v=\det (-\Delta )_{v,D}^{\frac 12}
{\det }' (-\Delta )_{v,N}^{\frac 12} , \quad
J_s=\det (-\Delta )_{s,D}^{\frac 12}
{\det }' (-\Delta )_{s,N}^{\frac 12} \eqno (20)$$
where the subscripts $v,s,D,N$ denote vectors, scalars, Dirichlet
and Neumann boundary conditions, respectively. The prime indicates
the exclusion of zero modes.

Let us change the variables in (20), $\{ v^a,s\} \to \{ u^a,s\}$,
$u^a=\epsilon^{ab}\Delta v^b-\nabla^bs$,
$$\cD v^a \cD s={\det }'(-\Delta )_{v,N} \cD u^a \cD s . \eqno (21)$$
By performing the Gaussian integration in (20) with the help of (21)
we obtain
$$Z=
{\det }' (-\Delta )^{-\frac 12}_{v,N}
{\det }' (-\Delta )^{-\frac 12}_{s,N}
{\det } (-\Delta )^{\frac 12}_{v,D}
{\det } (-\Delta )^{\frac 12}_{s,D} . \eqno (22)$$

Let us make use of the proper time representation of the path
integral
$$\log Z=-\frac 12 \int_0^\infty \frac {dt}t K(t) \eqno (23)$$
$$K(t)=K_{v,N}(t)+K_{s,N}(t)-K_{v,D}(t)-K_{s,D}(t),$$
where $K_{v,N}(t)={\rm tr}' \exp (t\Delta_{v,N})$ etc.
The corresponding heat kernels $K(t)$ can be evaluated with the
help of standard expressions [12]
$$K(t)=\frac 32 \sqrt{\frac {\pi}t} -3+O(t^{-\frac 12}).
\eqno (24)$$
The heat kernel (24) completely defines the ultra-violet divergencies
of $R^2+T^2$ gravity on the unit disk.

Now some comments are in order. We observe that the term with $t^{-1}$,
which is typical for two dimensions, is cancelled as it should be in
a model with equal number of bosonic (zweibein and connection) and
fermionic (ghosts) degrees of freedom. The surface divergence with
$t^{-1/2}$ is not cancelled. This is just a manifestation of the fact
that the ghosts obey Dirichlet boundary conditions while the gauge-fixed
fields satisfy the Neumann ones. Note, that if only a finite
number of modes is quantized the heat kernel expansion starts with
the zeroth power of the proper time $t$.

In conclusion, let us formulate the main lesson to be drawn  from
our present study. First, in the
presence of an independent connection field
it is possible to define local boundary conditions for fluctuations
of $\tilde e$ and $\omega$ in a diffeomorphism and local Lorentz
invariant way even in the case of a {\it not} totally geodesic boundary.
Second, these boundary conditions do not correspond to a quantum
integrable model with a finite number of modes. As a consequence,
a model of the latter type
 can be constructed either for more sophisticated
non-local boundary conditions or on a manifold with totally
geodesic boundary.

\vspace{5mm}

{\bf Acknowledgments}

\vspace{3mm}

This work was supported by the Fonds zur
F\"orderung der wissenschaftlichen Forschung, Project
P10221-PHY. The author is grateful to Wolfgang Kummer
for discussions, suggestions on the manuscript
and warm hospitality at the Technische
Universit\"at Wien.

\newpage

{\bf References}
\newline
1. G.Esposito, Quantum gravity, quantum cosmology and Lorentzian

geometries, Springer, Berlin, 1992.
\newline
2. K.Schleich, Phys. Rev. D32, 1889 (1985);

A.O.Barvinsky, A.Yu.Kamenschik and I.P.Karmazin, Ann. Phys. 219,

201 (1992);

A.O.Barvinsky, Phys. Rep. 230, 237 (1993).
\newline
3. I.G.Moss and S.Poletti, Nucl. Phys. B341, 155 (1990).
\newline
4. G.Esposito, A.Yu.Kamenschik, I.V.Mishakov and G.Pollifrone,
Phys.

Rev. D50, 6329 (1994).
\newline
5. A.O.Barvinsky, Phys. Lett. B195, 344 (1987).
\newline
6. M.O.Katanaev and I.V.Volovich, Phys. Lett. B175, 413 (1986);

Ann. Phys. 197, 1 (1990);

M.O.Katanaev, J. Math. Phys. 31, 882 (1990); 32, 2483 (1991).
\newline
7. W.Kummer and D.J.Schwarz, Phys. Rev. D45, 3629 (1992).
\newline
8. W.Kummer and D.J.Schwarz, Nucl. Phys. B382, 171 (1992);

F.Haider and W.Kummer, Int. J. Mod. Phys. A9, 207 (1994);

P.Schaller and T.Strobl, Class. Quantum Grav. 11, 331 (1993).
\newline
9. T.Chung and H.Verlinde, Nucl. Phys. B418, 305 (1994);

S.R.Das and S.Mikherji, Phys. Rev. D50, 930 (1994);

Mod. Phys. Lett. A9, 3105 (1994).
\newline
10. H.Luckock, J. Math. Phys. 32, 1755 (1991).
\newline
11. P.B.Gilkey, Invariance theory, the heat equation and the
Attiah-Singer

theorem, Publish or Perish, Delaware, 1984.
\newline
12. T.P.Branson and P.B.Gilkey, Commun. Part. Diff. Eqs. 15, 245
(1990)

\end{document}